\documentclass{aa}
\usepackage[varg]{txfonts}
\usepackage{lineno}
\usepackage{amsmath}
\usepackage{color}
\usepackage{multirow}

\titlerunning{VVV RR Lyrae type AB Classification}
\authorrunning{Elorrieta et al.}

\usepackage{comment}

\begin{document}
%\linenumbers

\title{A machine learned classifier for RR Lyrae in the VVV survey}

\author{Felipe Elorrieta\inst{1,2}\and
Susana Eyheramendy\inst{1,2}\and
Andr\'es Jord\'an\inst{3,2}\and
Istv\'an D\'ek\'any\inst{2,3}\and
M\'arcio Catelan\inst{3,2}\and
Rodolfo Angeloni\inst{4}\and
Javier Alonso-Garc\'ia\inst{5,2}\and
Rodrigo Contreras-Ramos\inst{2,3}\and
Felipe Gran\inst{3,2}\and
Gergely Hajdu\inst{3,2}\and
N\'estor Espinoza\inst{3,2}\and
Roberto K.\ Saito\inst{6}\and
Dante Minniti\inst{7,2,8}
}

\institute{Departmento de Estad\'istica, Facultad de
  Matem\'aticas, Pontificia Universidad
  Cat\'olica de Chile, Av.\ Vicu\~na Mackenna
    4860, 7820436 Macul, Santiago, Chile
\and Millennium Institute of Astrophysics, Santiago, Chile
\and Instituto de Astrof\'isica, Facultad de F\'isica, Pontificia Universidad
  Cat\'olica de Chile, Av.\ Vicu\~na Mackenna
    4860, 7820436 Macul, Santiago, Chile
\and Gemini Observatory, Chile
\and Unidad de Astronom\'ia, Facultad Cs. B\'asicas, Universidad de Antofagasta, Avda. U. de Antofagasta 02800, Antofagasta, Chile
\and Departamento de Física, Universidade Federal de Santa Catarina, Trindade 88040-900, Florianópolis, SC, Brazil
\and Departamento de Ciencias F\'isicas, Universidad Andres Bello, Rep\'ublica 220, Santiago, Chile
\and Vatican Observatory, V-00120 Vatican City State, Italy
}

\abstract{
Variable stars of RR Lyrae type are a prime tool to obtain distances to old stellar populations in the Milky Way, and one of the main aims of the Vista Variables in the Via Lactea (VVV) near-infrared survey is to use them to map the structure of the Galactic Bulge. Due to the large number of expected sources, this requires an automated mechanism for selecting RR Lyrae,and particularly those of the more easily recognized type $ab$ (i.e., fundamental-mode pulsators), from the $10^6-10^7$ variables expected in the VVV survey area. In this work we describe a supervised machine-learned classifier constructed for assigning a score to a $K_s$-band VVV light curve that indicates its likelihood of being $ab$-type RR Lyrae. We describe the key steps in the construction of the classifier, which were the choice of features, training set, selection of aperture and family of classifiers. We find that the AdaBoost family of classifiers give consistently the best performance for our problem, and obtain a classifier based on the AdaBoost algorithm that achieves a harmonic mean between false positives and false negatives of $\approx 7\%$ for typical VVV light curve sets. This performance is estimated using cross-validation and through the comparison to two independent datasets that were classified by human experts.  
}


\maketitle

%%%%%%%%%%%%%%%%%%%%%%%%%%%%%%%%%
%
% Introduction
% 
%%%%%%%%%%%%%%%%%%%%%%%%%%%%%%%%%%%

\section{Introduction}
\label{sec:intro}

Variable stars have been historically a prime tool for determining the content and structure of stellar systems and have had a crucial role in the history of Astronomy. The {\em General Catalog of Variable Stars} \citep{gcvs} lists over 110 classes and subclasses based on a variety of criteria. The broader distinction that can be made is between variables whose variation is intrinsic or extrinsic, depending on whether the process creating the observed variability is inherent to the star or not, respectively. Among the intrinsic variables, arguably the most important group is that of pulsating variables, as it contains the classes RR Lyrae and Cepheids, which satisfy a relation between their periods and their absolute luminosities, allowing the estimation of distances, a quantity both fundamental and elusive in Astronomy. For a current review of the physics and phenomenology of pulsating stars the reader is referred to the recent monograph by \citet{catelan:2015:book}

 RR Lyrae stars are of special importance for the purposes of exploring the distances and properties of old stellar populations. They have periods in the range $\approx$ 0.2-1 day, and are found only in populations that contain an old stellar component with an age $\gtrsim$ 10 Gyr. Based on their light curve shapes, RR Lyrae were originally separated into subclasses $a$, $b$ and $c$, also known as their Bayley type \citep{bailey:1902}. It was later recognized that classes $a$ and $b$ pulsate in the fundamental radial mode, whereas subclass $c$ does so in the first overtone radial mode \citep{schwarzschild:1940}, and therefore modern usage distinguishes only between RR Lyrae types $ab$ and $c$ (sometimes alternatively termed RR0 and RR1, respectively), and additional (but much less frequent) sub-classes have been introduced for other pulsational modes.

 Arguably the most important property of RR Lyrae is that they can be used as standard candles, and they have played a prominent role in determining the 3-dimensional structure of our Galaxy ever since Shapley used them to determine the distance of various globular clusters and used that fact to determine the distance to the Galactic center \citep{shapley:1918}. An obvious use of RR Lyrae would be then to map the structure of the Galactic bulge, but that is an observationally challenging task due to the large extinction present and the large area to be covered. Providing such a 3-dimensional mapping of the bulge using RR Lyrae was one of the main motivations for the Vista Variables in the Via Lactea (VVV) ESO Public Survey \citep{minniti:2010:vvv}, which covered $\approx$ 520 deg$^2$ of the Galactic bulge and disk with the VISTA telescope in Paranal. The VVV will catalog $\approx 10^9$ sources in the $ZYJHK_s$ bandpasses, with variability of the bulge probed with order $\sim 100$ epochs in $K_s$ over a total timespan of $>5$~yr.

With an expected yield of $10^6-10^7$ variable stars over its whole footprint, human classification is not a viable path to identify the different variability classes that will arise from the VVV. Machine learned procedures that classify variable stars, hereon the ``classifier'', become a must given these numbers and will become even more so as the size of future synoptic surveys such as the Large Scale Synoptic Telescope \citep[LSST,][]{lsst} become operational. The classifier receives as a  minimum input a time series of measured luminosity at irregularly sampled times, and outputs a score of confidence of membership in a specific class.

The aim of this study is to build an automated procedure to classify RR~Lyrae type $ab$ (from now on RR$ab$) stars from the VVV survey. The performance of automated classifiers of variable sources in the optical has been assessed in several previous studies \citep[e.g.][]{deb:2007:class,dubath:2011, richards:2011:class,paegert:2014, kim:2016}, which follow a similar approach. First one needs to assess if the variable is periodic, and if so, the period of the time series needs to be estimated. This is a crucial and challenging step as the astronomical time series are measured irregularly over time, often with intervals of high and low cadence. A parametric model is then fit to the folded light-curves. From the parameters of the model and directly from the light-curves a set of characteristics or {\em features} are then extracted. The final step is to decide on a specific function of the features that estimates the membership scores, and this is done for supervised classifiers such as the ones mentioned by using a {\em training} set of light-curves that have been labeled by a human expert. 

The near infrared (IR) offers at least two challenges as compared to the optical. First, it is intrinsically harder to classify RR$ab$ in the infrared due to the fact that their amplitudes are smaller than in the optical, making them harder to detect at a given signal-to-noise level, and the light curves are more sinusoidal, which increases the risk of confusion with other type of variables, particularly close binaries. Second, as opposed to the visible in which there are many high-quality light-curves with which supervised classifiers can be trained, in the near-IR high-quality templates are scarce, a situation that is now changing thanks to efforts such as the VVV Templates project \citep{vvvtemplates:2014}.

The structure of this paper is as follows. In \S~\ref{sec:data} we describe the data we use, in \S~\ref{sec:methods} we describe the methodology for feature extraction and the classification algorithm and its formal performance, and in \S~\ref{sec:performance} we further illustrate its performance by  comparing with a human expert classifier on various datasets extracted from the VVV. We close in \S~\ref{sec:conclusions} with a summary and our conclusions.

\section{Data}
\label{sec:data}

\subsection{The VVV ESO Public Survey}
\label{sec:vvv}

The Vista Variables in the Via Lactea (VVV) is an ESO public survey that is performing a variability survey of the Galactic bulge and part
of the inner disk using ESO's Visible and Infrared Survey Telescope
for Astronomy (VISTA). The survey covers $520$ deg$^2$ in the
$ZYJHK_S$ filters, for a total observing time of $\approx 1900$ hours, including $\approx 10^9$ point sources and an estimated $\sim 10^6-10^7$ variable stars. The final products will be a deep IR atlas in 5 passbands. One of the main goals is to gain insight into the inner Milky Way's
origin, structure, and evolution. This will be achieved by, for instance, obtaining a precise 3-dimensional map of the Galactic bulge. To achieve this goal, of particular importance are the RR Lyrae stars. There are many RR Lyrae in the direction of the bulge and, being very old, they are fossil records of the formation history of the Milky way. For a detailed account of the VVV see \citet{minniti:2010:vvv} and for a recent status updated with emphasis on variability see \citet{catelan:2014:vvv_update}

\subsection{Light Curve Extraction and Pre-processing}

Aperture photometry of VVV sources is performed on single detector frame stacks provided by the VISTA Data Flow System \citep{casu:2004} of the Cambridge Astronomy Survey Unit (CASU). A series of flux corrected circular apertures are used as detailed in previous publications \citep{catelan:2013, dekany:2015}. We denote the apertures as $\{1,2,3,4,5\}$ and are extracted in aperture radii of $\{0.5, 1/\sqrt{2}, 1, \sqrt{2}, 2\}$ arcsec.

Once $K_s$-band light-curves are extracted, stars with putative light variations were selected using Stetson’s J index \citep[][ Eq. 1]{stetson:1996}, taking advantage of the correlated sampling of the $K_s$ data, i.e., the fact that the light-curves are sampled in batches of 2-6 points measured almost at the same instance. We estimated significance levels of this statistic as a function of the number of points by Monte Carlo simulations using Gaussian noise, and selected
objects showing correlated light variations above the 99.9\% confidence level.

After selecting the curves that show evidence of variability, we proceed to eliminate individual observations that have anomalously large error bar estimates, as these indicate epochs with anomalous observing conditions which provide little information. If $\{\sigma_i\}_{i=1}^n$ is the set of uncertainty estimates for a light curve with $n$ points, we eliminate any points whose $\sigma_i$ values are $>5\sigma$ from the distribution mean, where $\sigma \equiv Var(\{\sigma_i\})$.

\subsection{Training sets}
\label{ssec:training_sets}

In order to use a supervised classification scheme, we need a training set that is used by the classifier to learn how RR Lyrae are characterized in feature space. In this paper, we decided to restrict ourselves to the $ab$-type RR Lyrae stars, as RR$c$ stars have smaller amplitudes (hence noisier light-curves) and are frequently very difficult to set apart from contact eclipsing binaries.
The training set are known instances of the RR$ab$ class, ideally observed with a cadence and precision similar to that of the target data that arises from the VVV. To retrieve a training set from VVV itself, we use light curves consistent with being variable from the bulge fields B293, B294, B295, located around Baade's window. These fields are well covered by OGLE-III \citep{ogle3}, and given that the OGLE-III $I$-band catalog is deeper than what we achieve with VVV in this region of the bulge, we make the assumption that all RR$ab$ in the three chosen fields that can be detected with VVV are present in the OGLE-III catalog. Therefore, we assemble a training set for each of the bulge fields by cross-matching the RR$ab$ catalogued in OGLE-III with the light curves extracted from VVV data.

In addition to the training set above, we make use of $K_s$ IR light-curves gathered as part of the VVV Templates project \citep{vvvtemplates:2014}. This project is a large observational effort aimed at creating the first comprehensive  database on stellar variability in the near-IR, producing well-defined, high-quality, near-IR light-curves for variable stars belonging to different variability classes, such as RR Lyrae, Cepheids, Eclipsing Binaries, etc. The main goal of the project is to serve as a training set  for the automated classification of VVV light-curves.

Table~$1$ shows the numbers of RR$ab$ light curves versus those belonging to other classes in each of the training datasets considered in this study. We note that while in principle we have color information for our objects, but similarly to other studies \citep[e.g.][]{kim:2016} we have decided not to use that information to train the classifier. The reason is that the footprint of the VVV is subject to a very wide range of extinction values \citep[e.g.,][]{gonzalez:2012}, and thus usage of color information would imply the need to account for the redenning, introducing another level of complexity.

\begin{table}
\caption{Number of RR$ab$ versus other classes in the training datasets.} \label{tab:totalStars}
\centering
\begin{tabular}{ccc}

\hline \hline
Dataset & Class & $N$\\
\hline
VVV Templates & RR$ab$  & 1603\\
& Other  & 1063\\ \hline
B293 Field & RR$ab$  & 277\\
 & Other   & 4869\\ \hline
B294 Field & RR$ab$  & 207\\
 & Other   & 5448\\ \hline
B295 Field  & RR$ab$  & 178\\
 & Other   & 4056\\
 \hline
 \end{tabular}
 \end{table}

\subsection{Measuring Classifier Performance}
\label{ssec:data_perf}

To assess the performance of the classifiers, using 10-fold cross validation we estimate four measures of quality: precision, recall, $F_1$ and AUC (defined below). In 10-fold cross-validation, the classifier is trained with nine tenths of the training set and with the remaining tenth, the performance of the classifier is assessed. This is done ten times, and each time a different tenth of the training set is held out, obtaining ten estimates of performance which are then averaged.

\begin{itemize}
\item {\em Precision}: is the probability that a randomly selected object predicted to be in a target class does belong to the target class. We denote precision by $P$. We note that the false discovery rate, i.e. the rate of false positives, is $1-P$.
\item {\em Recall}: is the probability that a randomly selected objected belonging to a target class is indeed predicted to be in that class. We denote recall as $R$.
\item $F_1$ measure, defined as the harmonic mean between $P$ and $R$, i.e. $F_1 \triangleq \frac{2PR}{(P+R)}$.  As a weighted average of precision and recall, it is a measure of the accuracy of the classifier, where a perfect accuracy would imply values close to one.
\item Area under the Curve (AUC): this is the area under the so-called receiver operating characteristic (ROC) curve, which shows the true-positive rate as a function of the false-positive rate. Values close to 1, which is the maximum possible, are best, as that indicates a classifier that quickly achieves a large true-positive rate with a correspondingly low false-positive one.

\end{itemize}

In order to evaluate these metrics, the continuous class {\em scores} returned by the classifier have to be transformed into class membership. In order to do so, we define a threshold score $p_t$, and classify all sources with scores $p > p_t$ as RR$ab$ and those below as ``other". We choose $p_t$ as the threshold that maximizes the $F_1$ metric estimated using 10-fold cross-validation of the training set; {\em all performance metrics presented in this work are calculated using a threshold determined using this rule}.

In addition to testing the performance using cross-validation, we use distinct datasets extracted from the VVV which were not used at all in the training and in which a catalog of RR$ab$ had been built by a human expert. We use two such datasets: (1) a catalog of RR$ab$ in the Galactic globular clusters 2MASS-GC 02 and Terzan~10 \citep{alonso:2015}. These RR$ab$ were classified based on their light curve shape and their position in the HR diagram; (2) a catalog of RR$ab$ in the outer bulge area of the VVV \citep[][the outer bulge refers to VVV fields with $b<-8$ deg]{gran:2016}. This is an extension of previous work reporting the RR$ab$ content of one single tile of the VVV \citep{gran:2015}. The RR$ab$ in this work were classified solely based on their shape by a human expert. %Compar

%%%%%%%%%%%%%%%%%%%%%%%%%%%%%%%%%
%
% Methods
% 
%%%%%%%%%%%%%%%%%%%%%%%%%%%%%%%%%%%

\section{Methods}
\label{sec:methods}

\subsection{Period estimation}
\label{ssec:period}

We estimate the periods of the light-curves using the Generalized Lomb-Scargle  periodogram \citep[GLS][]{zechmeister:2009}. We restrict the periodogram to frequencies satisfying $f_1 < 5$ day$^{-1}$.
We also eliminate all light curves whose highest Lomb-Scargle peak has a value $\leq 0.3$. The threshold was determined using the Lomb-Scargle peaks of known RR$ab$ in our VVV training fields, no source has a Lomb-Scargle peak with a value lower that the chosen threshold. Finally, we eliminate all curves with $\leq 50$ observations\footnote{VVV observations are clustered, usually in groups of $\approx$4, so the chosen cut corresponds to cutting light curves with typically $\lesssim 15$ epochs.}.

\subsection{Harmonic Model Light Curve Fit}
\label{ssec:harmonic_fit}

Once we have determined the first frequency we fit a harmonic model to the light curves.  Let $y(t)$ be the observed  light intensity at time $t$ from a given variable star,  let $\hat{y}(t)=a+bt$ be the linear trend estimated from a linear regression model of the photometric time series and let $r(t)=y(t)-\hat{y}(t)$ be the photometric time series with the linear trend subtracted.  Then we iterate between two steps: 

\begin{enumerate}
\item Perform a Fourier analysis to $r(t)$ to determine additional periodicities that might exist using GLS. The periodogram is calculated and the highest peak is selected. The corresponding frequency $f$ is used to find the parameters of the following harmonic fit using  the method of weighted least-squares estimation using the inverse measurement variances $\sigma_i^{-2}$ as weights:

\begin{equation}
\hat{z}(t)=\sum_{j=1}^m(a_j\sin(2\pi f j t)+b_j\cos(2\pi f j t))+b_0.
\end{equation}

When this model is fit using the first frequency $f_1$ an outlier rejection is performed (see below). This is not done for subsequent frequencies.

\item Reassign $r(t)$ by doing $r(t) \leftarrow r(t) - \hat{z}(t)$

\end{enumerate}

In words, we first subtract the linear trend from the photometric time series. Then, using the periodogram we identify the largest peak and use the corresponding frequency to fit a harmonic model with $m$ components. This new curve, together with the linear trend, are subtracted from the photometric time series and a new frequency is searched for in the residuals using the periodogram.  The new frequency is used to fit a new harmonic model. This process continues until $n$ frequencies are found and $n$ harmonic models with $m$ components are estimated. 

Finally, the $n$ frequencies are used to make an harmonic best-fit to the original light curve via weighted least squares of the full model given by

\begin{equation}
\hat{y}(t)=\sum_{i=1}^n\sum_{j=1}^m(a_{ij}\sin(2\pi f_i j t)+b_{ij}\cos(2\pi f_i j t))+a+bt.
\end{equation}

Following \citet{deb:2007:class} we use $n=3$ frequencies and $m=4$ harmonics to characterize the light curves.

\subsection{Outlier and Poor Light-Curve Rejection}
\label{ssec:outlier}

After finding the first frequency $f_1$ and fitting an harmonic model with $f_1$ only, we reject outliers from each light curve. 
We obtain a smooth estimate of the phased light curve using  smoothing splines obtained from the \texttt{R} function \texttt{smooth.spline}, with the parameter that controls the smoothing set to \texttt{spar=0.8} (this value was chosen  based on  a best fit measure of folded light curves). Then, we perform an iterative $\sigma$-clipping procedure to the residuals around the smooth model of the phased light curve. Assuming Gaussian errors, we remove outliers at the 4$\sigma$ level or above, and estimate the dispersion of the residuals robustly by setting $\sigma=1.4826\times$MAD, where MAD is the median absolute deviation. 

We also eliminate from our sample light-curves which have either too low signal-to-noise ratio or that have patchy phase coverage. In detail, we eliminate all light curves whose scatter around the phased light curve is not significantly different from the raw light curve by eliminating all curves whose median absolute deviation about the phased light curve is $>0.8$ times the median absolute deviation of the raw light curve, or in other words, light curves whose scatter is not significantly reduced after folding with the period\footnote{This corresponds in terms of the feature p2p\_scatter\_pfold\_over\_mad to 1/p2p\_scatter\_pfold\_over\_mad $> 0.8$}. To eliminate curves with incomplete phase coverage, we eliminate all curves where $1-\Delta\phi_{\rm max} < 0.8$, where $\Delta\phi_{\rm max}$ is the maximum of the consecutive phase differences $\{\phi_{i+1}-\phi_i\}_{i=1}^N$, where $N$ is the number of measurements and we take $\phi_{N+1}-\phi_N \equiv 1+\phi_1-\phi_N$.

\subsection{Feature extraction}
\label{ssec:features}

Features are extracted both from the raw and  phase-folded light curves, and from the parameters of the best fitting harmonic model. We adopt most of the features proposed in \citet{deb:2007:class}, \citet{richards:2011:class} and \citet{richards:2012:active}, and we introduce a few new features specifically designed to better discriminate RR$ab$ from close binaries such as those of the W~UMa type which turn out to be our most troublesome contaminant. Table~\ref{tab:features} lists all the features used by our classifier, along with a small description and a reference to the literature when adopted from previous work.  A total of 68 features are extracted to be used by the classifiers.

\begin{table*}
\caption{List of light-curve features used in this work}\label{tab:features}
\centering
\begin{tabular}{ccc}

\hline\hline
Feature name & Description\tablefootmark{a} & Reference\tablefootmark{b}\\
\hline
intercept (slope) & Intercept (slope) of a linear regression to the light curve & D07\\
$A_{ij}$ & Amplitude of the $i$-th frequency and $j$-th harmonic & D07 \\
$\phi_{ij}$ & Phase of the $i$-th frequency and $j$-th harmonic &  D07\\
$f_i$ & $i$-th frequency obtained from GLS & D07\\
$P_i$ & Peak in the GLS periodogram of the $i$-th frequency & D07\\
var$_i$ & Variance left after $i$-th fit of Fourier model &  D07\\
mse$_i$ & Mean Squared Error of $i$-th fit of Fourier model & D07\\
skew & Skewness of $y$ & R11 \\
small\_kurtosis & Small sample kurtosis of $y$ & R11\\
std & Standard deviation of $y$  & R11\\
max\_slope &  max$\{(y_{i+1}-y_i)/(t_{i+1}-t_i)\}$ & R11\\
amplitude & $\text{max}(y) - \text{min}(y)$ & R11\\
median\_absolute\_deviation & Median absolute deviation (MAD) of $y$ & R11\\
median\_buffer\_range\_percentage & Fraction of points in $\{y\}$ with amplitude within $<0.1$ of median$(y)$ & R11\\
pair\_slope\_trend & For the set $\{y_{N-29+i}-y_{N-30+i}\}_{i=2}^{30}$ the ratio $N_+/N_-$ & R11\\
flux\_percentile\_ratio\_mid\_k & $F_{50-k/2,50+k/2}/F_{5,95}$ & R11\\
percent\_amplitude & max$(|F-\text{median}(F)|)/\text{median}(F)$ & R11\\
percent\_difference\_flux\_percentile & $F_{5,95}/\text{median}(F)$ & R11\\
freq\_amplitude\_ratio\_21 (31) & Amplitude ratio of 2nd (3rd) to 1st component of the Fourier model & R12\\
freq\_frequency\_ratio\_21 (31) & Frequency ratio of 2nd (3rd) to 1st component of the Fourier model & R12\\
freq\_model\_max(min)\_delta\_mags & Difference in magnitudes between the two  maxima (minima) of $y_{2P}$ & R12\\
freq\_model\_phi1\_phi2 &  ($\phi_{min,1}-\phi_{max,1}) / (\phi_{min,1}-\phi_{max,2})$ (for $y_{m,2P}$) & R12\\
freq\_rrd & Boolean that is 1 if freq\_frequency\_ratio\_21 (or 31) is within $0.0035$ of 0.746 & R12\\
gskew & $(\text{median}(y)-\text{median}(y_{0})) + (\text{median}(y)-\text{median}(y_{1-p}))$ with $p=0.03$ & R12 \\
scatter\_res\_raw & $\text{MAD}(y-y_m) / \text{MAD}(y)$  & D11\\
p2p\_scatter\_2praw & $\sum_{i=2}^N(y_{2P,i+1}-y_{2P,i})^2 / \sum_{i=2}^N(y_{i+1}-y_{i})^2$ & D11\\
p2p\_scatter\_over\_mad & $\sum_{i=2}^N|y_{i+1}-y_{i}|  / (N-1)\text{MAD}(y)$ & D11\\
p2p\_scatter\_pfold\_over\_mad & $ \sum_{i=2}^N|y_{P,i+1}-y_{P,i}|/ (N-1)\text{MAD}(y)$  & D11\\
medperc90\_2p\_p & 90th percentile of $\{y-y_{m,2P}\}$ / 90th percentile of $\{y-y_{m,P}\}$& D11\\
fold2P\_slope\_10percentile (90) & 10th (90th) percentile of slopes $y_2P$ & R12\\
R1\tablefootmark{c} &  $(\phi_{max,1}-\phi_{min,1}) / (\phi_{min,1}-\phi_{max,2})$ (for $y_{s,2P}$) & R12, this work\\
R2 & $(y_{s,2P}(\phi_{max,1})-y_{s,2P}(\phi_{min,1})) / (y_{s,2P}(\phi_{min,1})-y_{s,2P}(\phi_{max,1}))$ & This work\\
\hline
\end{tabular}
\tablefoot{
\tablefoottext{a}{We use the following notation: the light curve magnitudes at times $t_i$ are denoted by $y(t_i)$ or $y_i$, the magnitudes phased with period $P$ at phase $\phi_i$ as $y_P(\phi)$, the harmonic (Fourier) model as $y_m$, the smooth spline mode as $y_s$. $\phi_{max(min),i}$ denotes the phase corresponding to the $i$-th maximum (minimum) of a phased curve,
$y(\phi_{max(min),i})$
the corresponding value. $N_+$ and $N_-$ denote the number of positive and negative members of a set, respectively. $F_{a,b}$ is the difference in flux between the percentile $a$ and $b$ of the fluxes implied by $y$. $y_{a:b}$ are the subset of $y$ whose members lie between the $a$-th and $b$-th percentile.
}
\tablefoottext{b}{D07=\citet{deb:2007:class}; R11=\citet{richards:2011:class}; R12=\citet{richards:2012:asas}; D11=\citet{dubath:2011}}
\tablefoottext{c}{This feature is the same as freq\_model\_phi1\_phi2 but uses $y_s$ instead of $y_m$.}
}
\end{table*}

An important set of features is derived from the harmonic fit, but care must be taken to choose parameter expressions which are invariant to time translations.  The frequencies $f_i$, together with the Fourier parameters $a_{ij}$ and $b_{ij}$, constitute  the direct set of parameters with which we model  the light curves. A drawback of this representation is that the parameters are not invariant to time translation. In other words, if from the same star there are two light curves observed which do not coincide in its starting time, then these two light curves will have two different sets of parameters to represent the same star.  To obtain parameters that uniquely represent a light curve we transformed the Fourier coefficients into a set of amplitudes $A_{ij}$ and phases $\phi_{ij}^{'}$ as follows:

\begin{equation}
A_{ij}=\sqrt{a_{ij}^2+b_{ij}^2},
\end{equation}

\begin{equation}
\phi_{ij}^{'}=\arctan(\sin(\phi_{ij}),\cos(\phi_{ij}))
\end{equation}

\noindent where,

\begin{equation}
\phi_{ij}=\arctan(b_{ij},a_{ij}) -\frac{jf_i}{f_1}\arctan(b_{11},a_{11}).
\end{equation}

Note that $\phi_{11}$ is chosen as the reference and is set to zero and that $\phi^{'}_{ij}$ takes values in the interval $[-\pi,\pi]$.

\begin{figure*}[ht!]
\centering
\includegraphics[width=0.75\textwidth]{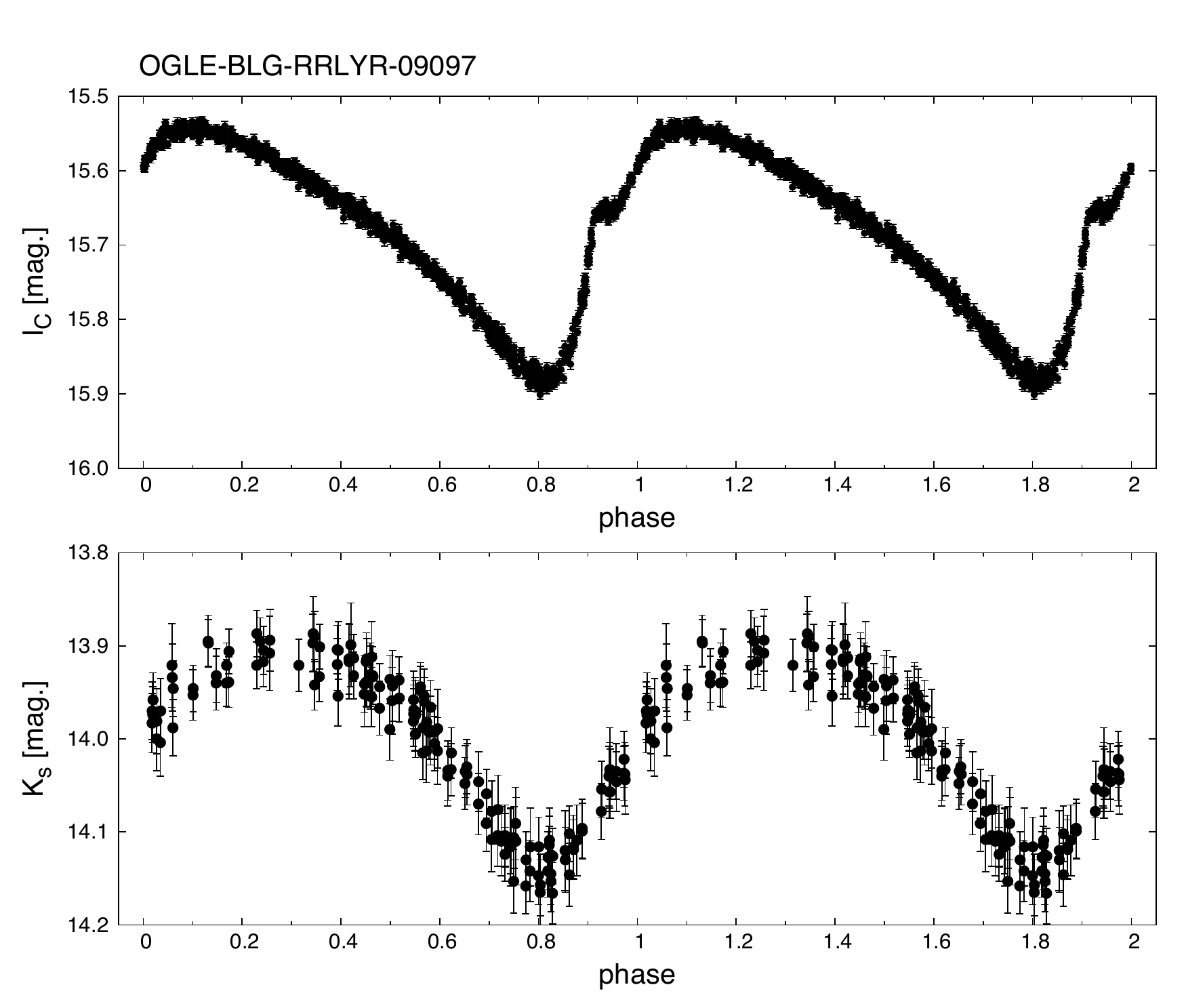}
\caption{Example of a known RR$ab$ classified by OGLE using an optical $I_C$ light curve (upper panel) and that shows a very symmetric light curve in the infrared (lower panel, $K_s$ light curve from the VVV).}
\label{fig:troll}
\end{figure*}

We used two additional features in this work which are listed at the end of Table~\ref{tab:features}. As mentioned in \S~\ref{sec:intro}, the RR$ab$ light curves in the IR are of lower amplitude and more symmetrical than in the optical. This makes confusion with close binaries that have periods in the range of RR$ab$ a very challenging contaminant for the classifier to discriminate. Even if subtler in the near-IR, one of the features that often distinguishes RR$ab$ is some level of asymmetry around the peak of the light-curve, with a stepper ascent and a slower descent. As we will see later, even though one of the  features we introduced has some importance, it is not a very large one. This is not surprising once one realizes that some {\em bona-fide} RR$ab$ are extremely symmetrical in the IR. Indeed, we show in Figure~\ref{fig:troll} the VVV ($K_s$) and OGLE ($I_C$) light curves of a known RR$ab$ classified by OGLE. The figure clearly shows the challenges of classifying RR$ab$ in the IR as compared to the optical. It is also clear from the figure that any feature that tries to quantify asymmetry in the peak cannot be very relevant for a curve such as this one and the classifiers learn this fact.

\subsection{Choice of Classifier}
\label{ssec:classifier}

 To choose the best classifier for RR$ab$ we test several well-established classifiers. We direct the reader to \citet[][]{hastie:2009} for a detailed description of many of the algorithms below (see also \citet{ivezic:book}); more detailed documentation can be found in the documentation of the functions we use. The classifiers have been implemented using functions in the \texttt{R} language \citep{Rpkg}; we indicate the functions used as well as the options/parameters used when appropriate.

 \begin{itemize}
 \item Logistic regression classifier. We used the \texttt{glm} function  of the \texttt{stats} library to perform the regression.
 \item Classification tree (CART). We used the \texttt{rpart} function of the  library with the same name.
 \item Random Forest (RF). We use the \texttt{randomForest} function of the \texttt{randomForest} library in R  with parameter \texttt{ntree=500} (number of trees) and \texttt{mtry=20} ( number of variables at each tree). To set these parameter values, we  assessed the performance of the classifier on a grid of values. \texttt{ntree} took values  in the interval \texttt{[100-1000]}, and  \texttt{mtry} took values in the interval \texttt{[1,p]}, where p is the number of features. Our final classifier includes only $12$ out of the $68$ original features;  for the final classifier the parameter \texttt{mtry} was set to $3$.
 \item Stochastic Boosting (SBoost),  is implemented  with the \texttt{ada} function of the \texttt{ada} library in R. Two versions of this classifier are tested. The first one has parameters \texttt{loss}=``exponential'' and parameter \texttt{type}=``discrete'' (denoted as Sboost$1$) and the second has parameters \texttt{loss}=``logistic'' and parameter \texttt{type}=``real'' (denoted as Sboost$2$), both with parameter  \texttt{max.iter}=$500$. All combinations of loss functions with the different options for the  \texttt{type} parameter were evaluated, and the \texttt{max.iter} parameter was assessed in the interval $[100,1000]$. Based on performance we chose to consider SBoost$_1$ only in what follows, and we will write just SBoost for short.
 \item AdaBoost (ADA, short for adaptive boosting) is implemented with the \texttt{boosting} function of the \texttt{adabag} library in R. Ada.M1 uses the AdaBoost.M1 algorithm \citep{freund:1996} with Breiman weight updating coefficient (option \texttt{coeflearn}=``Breiman"), while Ada.SAMME uses the AdaBoost SAMME \citep{zhu:2009} algorithm with option \texttt{coeflearn}=``Zhu". The  \texttt{coeflearn} values represents different options for weighting the ``weak" classifiers based on misclassification rate. In both classifiers we assessed the number of trees (parameter \texttt{mfinal}) in the interval $[100,1000]$, and based on cross-validation chose \texttt{ntrees}=500. Both algorithms have the parameter \texttt{boos} set to \texttt{TRUE}, and therefore, a bootstrap sample of the training set is drawn using the weights for each observation on that iteration. Based on performance we chose to consider Ada.M1 only in what follows.
 \item A Support Vector Machine (SVM) classifier with a polynomial kernel is implemented with the function \texttt{svm} of the \texttt{e1071} library with parameters \texttt{degree=2} and \texttt{nu=0.1}.  The choice of radial basis, lineal,  polynomial and sigmoid kernels were assesed, and we found that the best performing was the polynomial kernel. 
 %\item $k$-nearest neighbour (kNN, see \S~XX in HT09).
 \item LASSO is a penalized likelihood classifier that implements  $L_1$ penalization. Implementation was done via the \texttt{glmnet} function of the \texttt{glmnet} library with options \texttt{family}=``Binomial" and \texttt{nlambda}=1000. The latter was chosen after testing performance in the rante [100,10000]. The parameter $\alpha$, the elasticnet mixing parameter was tested in the range [0,1] and set to 1 (giving thus a LASSO, $\alpha=0$ corresponds to ridge regression).
 \item Multiple Hidden Neural Networks (MHNN) implemented with the \texttt{nn.train} function in the \texttt{deepnet} package in R, and parameters \texttt{hidden=10}, \texttt{activationfun}=``sigm" (a sigmoid activation function),   \texttt{batchsize}=1500 and \texttt{numepochs=2000}. The parameters were set after the classifier performance was assessed testing the parameters \texttt{batchsize} in the interval [100, 2000] and the number of hidden layers in [1, 20]. 
 \item Deep neural network (DeepNN) implemented with  the \texttt{dnn} function in the \texttt{deepnn} package in R. A sigmoid activation function and 5 hidden layer were useds, with parameter \texttt{batchsize}=1500, \texttt{numepochs=2000} and $10$ hidden layers. The classifier performance was assessed testing the parameters \texttt{batchsize} in the interval [100,2000] and the number of hidden layers in [1, 20].

 \end{itemize}

  To choose the best classifiers we estimate, using $10$-fold cross validation on the training set, the area under the curve (AUC), precision $P$, recall $R$ and the $F_1$ measure, all defined in \S~\ref{ssec:data_perf}. The training set was chosen to be the VVV templates plus 80\% field B293, 80\% field B294, and 80\% field B295 (we show below that this particular choice of training set is representative of the other fields). The cross-validation estimates of performance that results after training all of the classifiers listed above using all the features available are summarized in Table~\ref{tab:perf_CV}. It is clear from this table that when using all the features we have defined the AdaBoost and SBoost classifiers achieve best performance. It is interesting to note that the performance of AdaBoost and SBoost classifiers is significantly better than that of Random Forests, which has been the classifier of choice in the recent literature \citep[e.g.][]{dubath:2011,richards:2012:asas}. To the best of our knowledge this is the first time AdaBoost and SBoost have been applied to the problem of variable star classification, and based on our results we suggest it should always be tried. 
  While AdaBoost and SBoost are fairly equivalent within the uncertainties, we choose Ada.M1 as our final classifier. Of course, not all features are equally important for classification and we will re-assess in \S~\ref{ssec:feature_selection} below the relative performance of the classifiers we are considering when restricting to the set of features that capture the bulk of the problem.

\begin{table}

\caption{Cross-Validation Performance of Classifiers on the Templates+B293+B294+B295 Training Set, using all features} \label{tab:perf_CV}
\centering
\begin{tabular}{ccccc}

\hline\hline
Algorithm & AUC & $P$ & $R$
& $F_1$ ($\sigma_{F_1}$)\\
\hline
Logistic & 0.9756 & 0.7869 & 0.8579 & 0.8121 $(\pm 0.0198)$ \\
CART & 0.9265 & 0.8591 & 0.7373 & 0.7911 $(\pm 0.0177)$ \\
RF & 0.9811 & 0.9515 & 0.8234 & 0.8804 $(\pm 0.0105)$ \\ %3 mtry=3
SBoost & 0.9939 & 0.9522 & 0.9094 & 0.9298 $(\pm 0.0054)$ \\ %5 Exponential Loss Discrete Adaoost
Ada.M1 & 0.9937 & 0.9685 & 0.8974 & 0.9311 $(\pm 0.0046)$ \\ %7 Breiman
SVM & 0.9792 & 0.9036 & 0.7960 & 0.8456 $(\pm 0.0120)$ \\ %10 Kernel Polynomial
Lasso & 0.9849 & 0.8599 & 0.8398 & 0.8454 $(\pm 0.0139)$ \\
MHNN & 0.9851 & 0.9190 & 0.8793 & 0.8968 $(\pm 0.0116)$ \\
DeepNet & 0.9823 & 0.9143 & 0.8762 & 0.8941 $(\pm 0.0102)$ \\
\hline
\end{tabular}
\end{table}

\subsection{Choice of Aperture for Photometry}
 \label{ssec:aper}

After we select the best classifier, we assess whether the selection of an aperture from which we estimate the features of the light-curves has an effect on the classification performance. We implement three strategies for selecting the aperture. The first one is to fix the aperture size to be equal for all variable stars, and call this strategy \texttt{fixAper(i)} for aperture size $i$ (this gives us five strategies). Second, we choose for each light curve the aperture size that achieves the minimum sum of squared measurement errors, and call this strategy \texttt{minError}. Third, similarly to \citet{richards:2012:asas}, we develop a kernel density classifier that outputs probability scores for each aperture based on the median of the mean magnitudes at the five apertures. 

\begin{figure}
\centering
\includegraphics[width=0.5\textwidth]{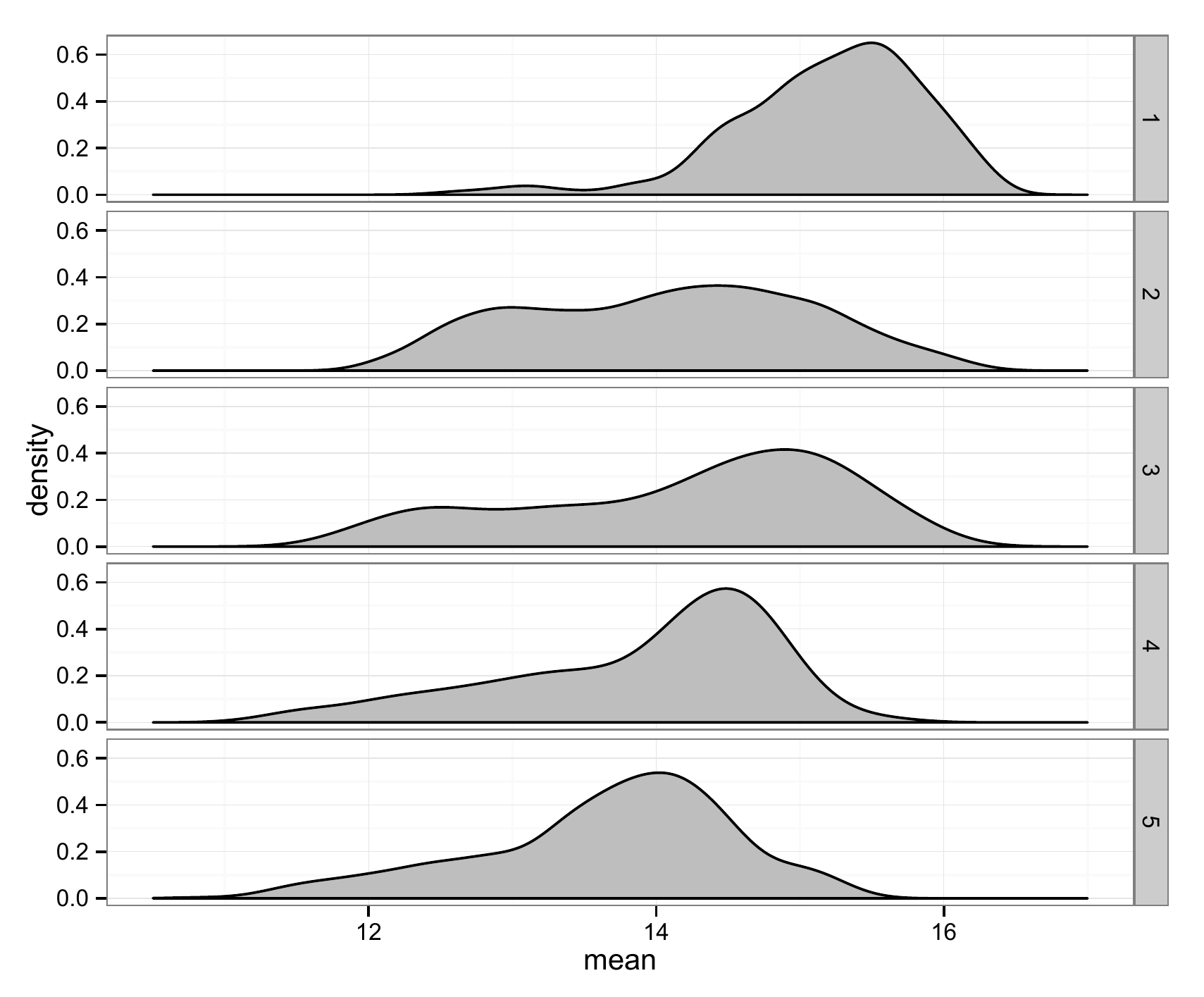}
\caption{Kernel density estimates of the mean magnitude of curves having their minimum sum of squared errors at each aperture size.\label{fig:kernelD}}
\end{figure}

\begin{figure}
\centering
\includegraphics[width=0.5\textwidth]{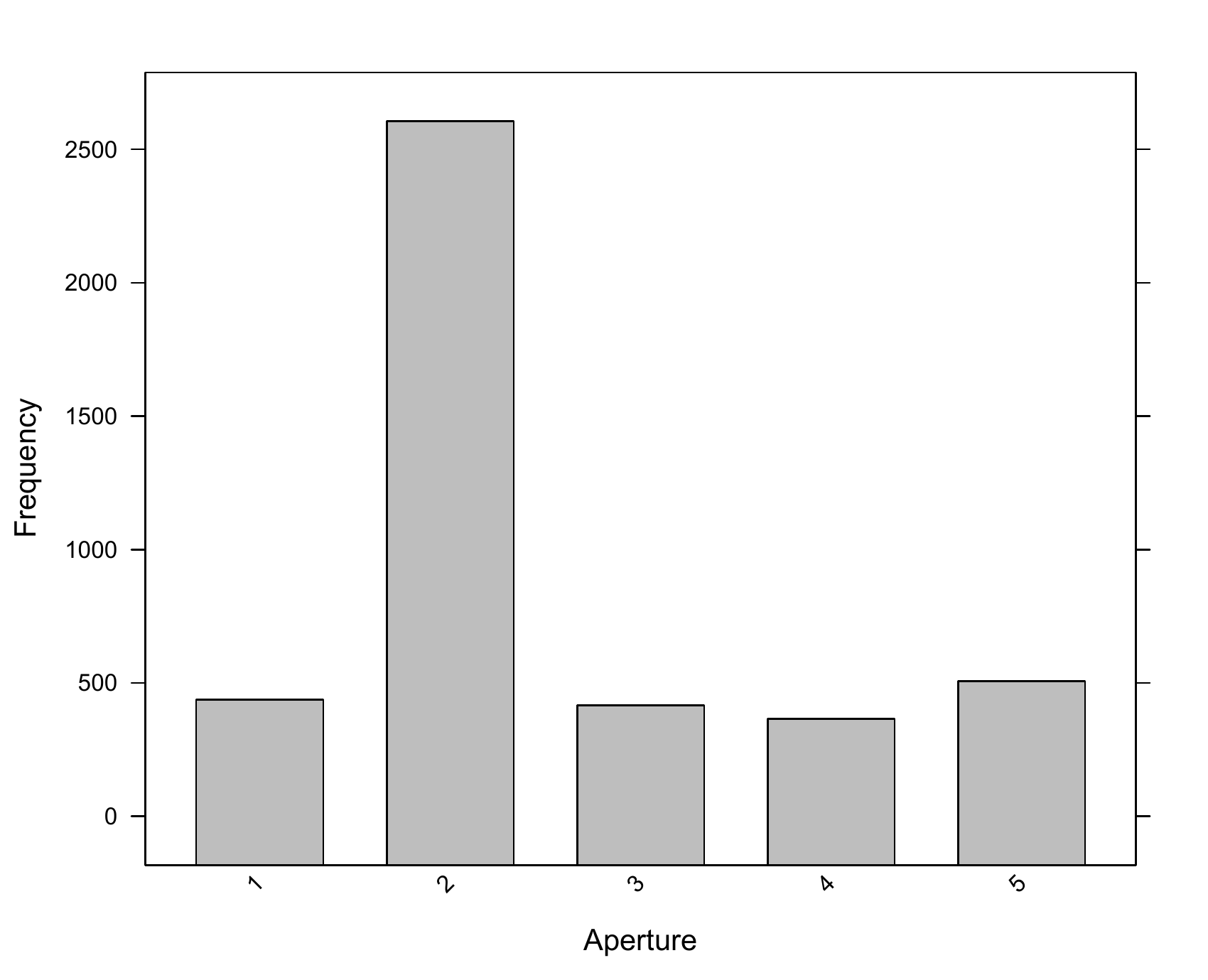}
\caption{Number of light-curves having their minimum sum of squared errors at each aperture size.\label{fig:apertureProp}}
\end{figure}

The kernel density for aperture $1$, say $g_1$, is estimated using the mean magnitudes of all the light-curves whose minimal sum of squared measurement errors is achieved at aperture $1$. In a similar way we estimate the kernel densities for the remaining four apertures, $g_2, g_3, g_4, g_5$. Figure~\ref{fig:kernelD} shows the densities for each aperture size. We also estimate the proportion of light-curves that achieve the minimum sum of squared measurement errors at each aperture, say $\pi_1, \pi_2, \pi_3, \pi_4, \pi_5$, estimates of which are shown in Figure~\ref{fig:apertureProp}. Unlike \citet{richards:2012:asas}, we develop a soft thresholding classifier to choose an optimal aperture.  For a new light-curve, we  choose the aperture whose probability $p(j|\mathbf{y})$ is the highest, where

\[
p(j|\mathbf{y})=\frac{\pi_jg_j(y^{\ast})}{\sum_{i=1}^5\pi_ig_i(y^{\ast})}
\]

\noindent and $y^{\ast}$ is the median of the mean magnitudes at all five apertures. The classifier selects the aperture with the highest score, we call this third strategy \texttt{KDC}. 

For each variable star then, we select the time series based on a particular strategy for selecting aperture and follow the full procedure of extracting the features and training the classifiers. We evaluate the performance of the seven strategies using the Ada.M1 classifier,  which we have determined already to be among the top performers. Results are shown in Table~\ref{tab:aperturePerformance}. We use cross-validation, on variable stars from the B295 field, to estimate the performance of the Ada.M1 classifier under different strategies. We can see from Table~\ref{tab:aperturePerformance} that the strategies KDC and \texttt{fixAper(2)} are the best performers. Given that most sources have their minimum sum of squared errors for aperture $i=2$ it is natural that KDC is equivalent in terms of performance to \texttt{fixAper(i)}, but for source magnitude distributions where this is not the case we would expect KDC to be better. We note that KDC outperforms \texttt{minError}, a strategy often used to choose an aperture.

\begin{comment}
\begin{table*}
\caption{$F_1$ Measure by Aperture and Classifier Algorithm} \label{tab:aperturePerformance}
\centering
\begin{tabular}{lcccc}

\hline\hline
Strategy & SBoost$_1$ & 
SBoost$_2$ & Ada.M1 & Ada.SAMME\\
\hline
fixAper(1) & 0.9210 (0.0189) & 0.9244 (0.0176) & 0.9120 (0.0165) & 0.9217 (0.0175) \\
fixAper(2) & 0.9359 (0.0090) & 0.9363 (0.0080) & 0.9308 (0.0099) & 0.9303 (0.0092) \\
fixAper(3) & 0.8794 (0.0197) & 0.8728 (0.0198) & 0.8799 (0.0271) & 0.8757 (0.0184) \\
fixAper(4) & 0.8621 (0.0181) & 0.8651 (0.0206) & 0.8615 (0.0207) & 0.8631 (0.0218) \\
fixAper(5) & 0.8055 (0.0249) & 0.8019 (0.0278) & 0.8146 (0.0194) & 0.8166 (0.0251) \\
KDC & 0.9316 (0.0090) & 0.9327 (0.0093) & 0.9308 (0.0099) & 0.9255 (0.0089) \\
minError & 0.9133 (0.0139) & 0.9140 (0.0136) & 0.9140 (0.0146) & 0.9131 (0.0159)\\
\hline
\end{tabular}
\end{table*}
\end{comment}

\begin{table}
\caption{$F_1$ Measure by Aperture Choice Strategy for Ada.M1} \label{tab:aperturePerformance}
\centering
\begin{tabular}{lcccc}

\hline\hline
Strategy &  Ada.M1 \\
\hline
fixAper(1) & 0.9120 (0.0165) \\
fixAper(2) & 0.9308 (0.0099) \\
fixAper(3) &  0.8799 (0.0271)  \\
fixAper(4) &  0.8615 (0.0207)  \\
fixAper(5) &  0.8146 (0.0194) \\
KDC &  0.9308 (0.0099)  \\
minError & 0.9140 (0.0146) \\
\hline
\end{tabular}
\end{table}

\subsection{Feature Selection}
\label{ssec:feature_selection}

\begin{figure}
\includegraphics[width=0.5\textwidth]{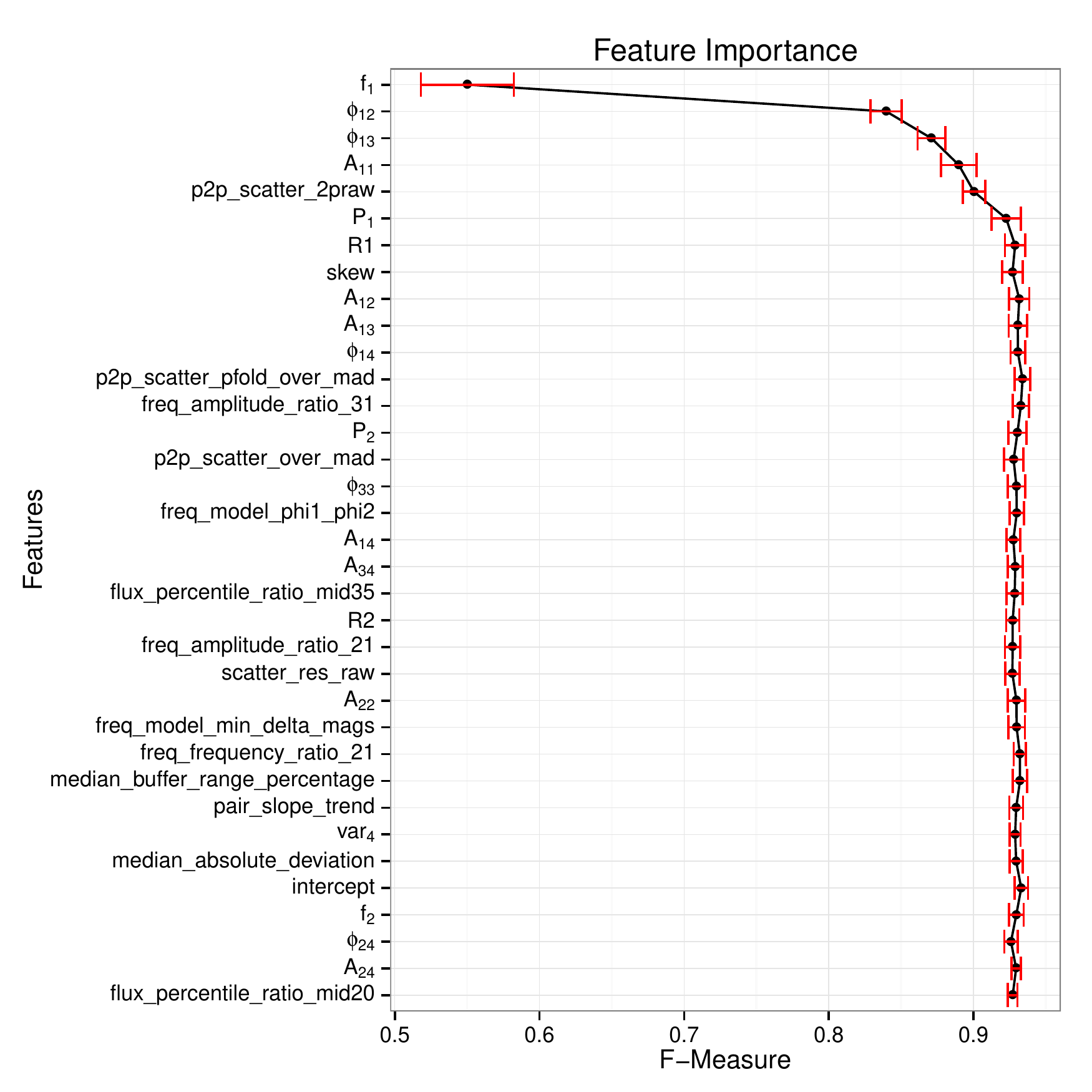}
\caption{Feature importance using the Ada.M1 classifier. Based on this graph we choose to consider only the $12$ most important features in the final classifier.}
\label{fig:feature_importance}
\end{figure}

In \S\ref{ssec:classifier} we used all 68 features to assess which is the better classifier. It is clear that not all features should have the same impact on the classification, and we can measure how important each one of them is for the classification. In Figure~\ref{fig:feature_importance} we show the features ordered by importance, with the most important on top. 
The relative importance of the predictor variables is computed using the gain of the Gini index given by a variable in a tree and the weight of the tree. Not surprisingly, by far the most important feature is the fundamental frequency ($f_1$). As stated in the Introduction, RR$ab$ have periods in a very well-defined range and then it is natural that these characteristics is the most prominent in deciding whether a curve belongs to the RR$ab$ class or not. It also underscores how important it is to estimate the period correctly, and indeed there have been many efforts in assessing what are the best tools for period finding in astronomical time series \citep[e.g.][]{graham:2013}. Therefore, obtaining accurate and robust periods is crucial for our classification scheme, if the period is not accurate the classification is bound to be imprecise. The next features in importance are all based on the harmonic fit to the data, and the first one unrelated to that is  \texttt{p2p\_scatter\_2praw}

Based on Figure~\ref{fig:feature_importance} we determined that using 12 features suffices to obtain the final best performance. While in principle one could keep all 68 features, doing so introduces a larger scope for noise being introduced by features that are not very informative and that particularly for faint sources can be noisy. The choice of 12 features is done on the basis that by the 12-th feature, $A_{12}$, the $F_1$ measure ($x$-axis in the Figure) has reached a maximum already and adding the next feature actually makes the performance decrease a bit, i.e. we have already reached at the 12-th feature the best performance over the level around which additional features do not add significant information. 

\begin{table}

\caption{Cross-Validation Performance of Classifiers on the Templates+B293+B294+B295 Training Set, using the best 12 features} \label{tab:perf_CV_12}
\centering
\begin{tabular}{ccccc}

\hline\hline
Algorithm & AUC & $P$ & $R$
& $F_1$ ($\sigma_{F_1}$)\\
\hline
Logistic & 0.8574 & 0.4624 & 0.8057 & 0.4855 $(\pm 0.0686)$ \\
CART & 0.9311 & 0.8577 & 0.7342 & 0.7860 $(\pm 0.0141)$ \\
RF & 0.9902 & 0.9522 & 0.8896 & 0.9194 $(\pm 0.0067)$ \\
SBoost & 0.9942 & 0.9483 & 0.9184 & 0.9326 $(\pm 0.0050)$ \\
Ada.M1 & 0.9937 & 0.9526 & 0.9154 & 0.9331 $(\pm 0.0055)$ \\
SVM & 0.9840 & 0.9073 & 0.8321 & 0.8651 $(\pm 0.0090)$ \\
Lasso & 0.9553 & 0.7618 & 0.6691 & 0.6953 $(\pm 0.0145)$ \\
MHNN & 0.9824 & 0.9365 & 0.8608 & 0.8956 $(\pm 0.0090)$ \\
DeepNet & 0.9823 & 0.9258 & 0.8775 & 0.9001 $(\pm 0.0079)$ \\
\hline
\end{tabular}
\end{table}

The cross-validation estimates of performance that results after training all of the classifiers using only the 12 more important features  are summarized in Table~\ref{tab:perf_CV_12}. It is interesting to note that the performance of the AdaBoost and SBoost classifiers is very similar to that when using all features, a desirable property as the performance is not overly sensitive to the particular choice of features. On the other hand, other classifiers change their performance significantly. Of particular interest is the fact that the performance of the widely used Random Forest classifiers comes closer to that of of AdaBoost and SBoost, although it is still lower. 

\subsection{Sensitivity to Training Set Choice}
\label{ssec:training_choice}

An important step on building a classifier is to select an appropriate training set that captures the variability of the data, as if the training set is not representative the resulting classifier is bound to fail for some kinds of objects. To test our sensitivity to the training set choice we train our classifier using different training sets by taking different subsets out of the four available sets (Templates, B293, B294, B295, see \S\ref{ssec:training_sets}). In Table~\ref{tab:sensitivity} we show the results of this exercise using the Adaboost.M1 classifier, showing the $F_1$ measure for different combinations of training and test sets; note that in this case we measure performance using a test set that is disjoint from the training.   The first row shows the performance for the classifier trained using only the Templates (T), the following three rows shows the performance for the classifier trained using the Templates plus  B295, B294 or B293 respectively. The next three rows are similar to the previous three ones, but now two complete fields are incorporated into the training set. The last two rows show the performance of the classifier with all of the curves from the Templates plus 90\% (80\%) of the curves from the three fields. As evident from Table~\ref{tab:sensitivity} the performance is best when we include curves from the Templates and all three fields. It does not vary significantly between having 80\% or 90\% of the curves over the expected random variations in the $F_1$ performance, which for Adaboost.M1 was expected to be on the order $1\%$.  
We conclude that our choice of training set of Templates+80\% B293 + 80\% B294 + 80\% B295 is not biasing our results in a significant way as assessed by training the classifier.

\begin{table}[ht]
\centering
\footnotesize{
\begin{tabular}{|r|rrr|}
 \hline
Training \textbackslash Test & B295 & B294 & B293\\ \hline
Templates & 0.8713 & 0.8905 & 0.9065 \\
T+B295  & - & 0.9095 & 0.9251 \\
T+B294   & 0.9043  & - & 0.9270 \\
T+B293  & 0.9003 & 0.9150 & - \\
All \textbackslash B294 & - & 0.9204 & - \\
All \textbackslash B293 & - & - & 0.9290 \\
All \textbackslash B295 & 0.9122 & - & -\\
All 90\% &  0.9267 & 0.9476 & 0.9502 \\
All 80\% &  0.9269 & 0.9536 & 0.9304 \\
 \hline
\end{tabular}
} 
\caption{F-Measure by Training Set (Adaboost.M1)} 
\label{tab:sensitivity} 
\end{table}

\subsection{Final Classifier}
\label{ssec:final_classifier}

Building upon our extensive experiments, as described above, we define our final classifier is the following way. We used Templates+80\% B293 + 80\% B294 + 80\% B295 as our training set, selected the aperture of each curve based on KDC, and adopted an Ada.M1 classifier using the following 12 features of the 68 listed in Table~\ref{tab:features} (ordered by importance)

\begin{enumerate}
\item $f_1$
\item $\phi_{12}$
\item $\phi_{13}$
\item $A_{11}$
\item \texttt{p2p\_scatter\_2praw}
 \item $P_1$
\item \texttt{R1}
\item \texttt{skew}
\item $A_{12}$
\item $A_{13}$
 \item $\phi_{14}$
  \item \texttt{p2p\_scatter\_pfold\_over\_mad}
\end{enumerate}

The $F_1$ measure as estimated from cross-validation is $\approx 0.93$. We will refer to this classifier in what follows as the final or ``optimal" classifier.
It is interesting to compare this performance with that obtained by similar methods in the optical for ground-based studies. The machine learned classifier implemented for ASAS by \citet{richards:2012:asas} attains an $F_1$ performance of $\approx 0.96$ for RR$ab$ classification (see their Figure 5), or about a factor of $\lesssim$2 better that what we get in terms of the expected number of false positives/negatives. The increased performance in the optical is expected given the larger amplitudes and more asymmetric shape in those bands, and thus the optical should be taken as an upper bound of what supervised classification could achieve for the VVV data. Of course the data are not necessarily directly comparable, and indeed the ASAS data used by \citet{richards:2012:asas} have a larger number of epochs (mean of 541, whereas the VVV has typically on the order of 100). We conclude from this comparison that our finally chosen performance is already fairly close to what we can think of as an upper bound for supervised classification methods for RR$ab$.

\section{Performance on Independent Datasets}
\label{sec:performance}

In the sections above we measured the performance of various classifiers, including what we choose as the optimal one. In this section we compare the performance of our classifier with the gold standard human expertise on two datasets mentioned in \ref{ssec:data_perf}: (1) a catalog of RR$ab$ in the Galactic globular clusters 2MASS-GC 02 and Terzan~10 \citep{alonso:2015}, and (2) a catalog of RR$ab$ in the outer bulge area of the VVV. This comparison is particularly relevant as it allows us to assess the generalization performance of the classifier on different datasets in which flux measurements do not necessarily follow the same conditions in cadence, depth, etc. as our training set. The cross-validation evaluation of the performance of our optimal classifier gives an $F_1$ measure of $\approx 93\%$ using a score threshold of 0.548, and so if that performance generalizes well we would expect the harmonic mean of the number of false positives and false negatives to be of order 7\%.

\subsection{RR$ab$ in 2MASS-GC 02 and Terzan~10}

\begin{figure}
\includegraphics[width=0.4\textwidth]{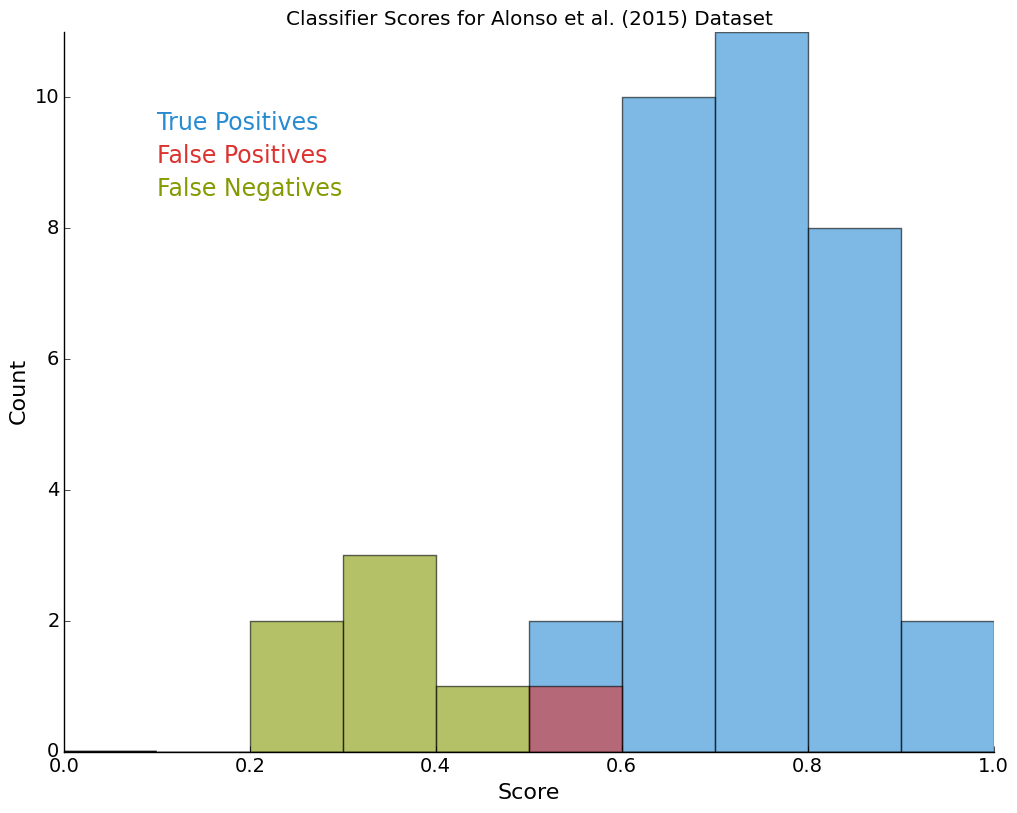}
\caption{Histogram of scores obtained by the classifier for the light curves of the \citet{alonso:2015} sample. Shown are the true positives (sources classified by \citet{alonso:2015} as RR$ab$), false positives and false negatives.}
\label{fig:cluster_scores_hist}
\end{figure}

In \citet{alonso:2015} 39 variables were classified as RR$ab$. In addition to the light curves, \citet{alonso:2015} used color information in order to assess the nature of the variables, in addition to optical light curves from OGLE when available. Therefore, there is high confidence on the nature of the stars classified as RR$ab$. Note that this comparison is  biased against our machine-learned classifier, which only had the VVV light curves as input for the classification. The distribution of scores for the RR$ab$ classified by \citet{alonso:2015} (true positives), and the false positives/negatives, are shown in Figure~\ref{fig:cluster_scores_hist}. As evident in the figure, the great majority of known RR$ab$ are classified correctly as such by the classifier. Six sources, or $\approx 15\%$ of the sample, are false negatives. The periods for those sources are consistent with those of RR$ab$, and given the fact that they are not symmetrical they were classified as RR$ab$ by \citet{alonso:2015}.

\begin{figure}
\includegraphics[width=0.45\textwidth]{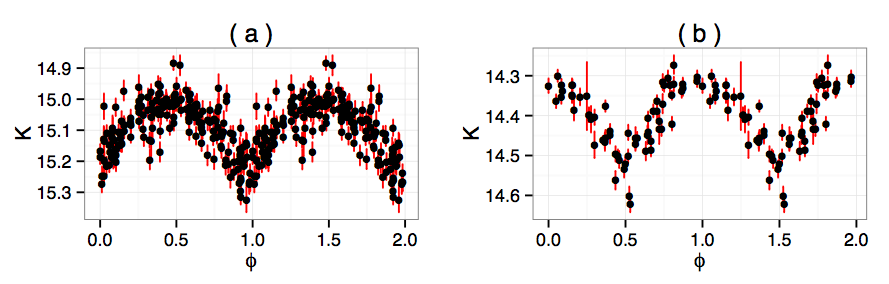}
\caption{Figure for two sources that were nominally false positives: (a) Terzan10\_V113; (b) internal identifier 273508. One of them (a) is a {\em bona fide} false positive, while the other (b) is actually a true positive that was not flagged as such in the work of \citet{alonso:2015} (see text) \label{fig:cluster_fp}
}
\end{figure}

On the other hand, there were formally two false positives, or $\approx 5\%$ of the sample at face value, which are shown in Figure~\ref{fig:cluster_fp}.  We discuss each in turn.  Terzan10\_V113, shown in panel (a), was classified as an eclipsing binary in \citet{alonso:2015} due to its very symmetric nature. It is also not classified as an RR$ab$ by OGLE, reinforcing its status as a non RR$ab$. In light of the very symmetric nature of some RR$ab$ in the IR (cf.\ Figure~\ref{fig:troll}) it is not surprising that variables with the right periods and amplitudes for RR$ab$ get classified as such even if they are very symmetric.
One additional variables was classified as RR$ab$ but was not present as a variable in the \citet{alonso:2015} catalog. Its internal IDs is 21089, and it is shown in panel (b) of Figure~\ref{fig:cluster_fp}. Variable 273508 is actually an RR$ab$ detected by OGLE (OGLE\_BLG\_RRLYR-33508) that was inadvertedly left out in \citet{alonso:2015}. It is most likely a field RR$ab$ as it is beyond the tidal radius of Terzan~10, and also it is too bright to be part of the cluster. In summary, after looking in detail at the formal false positives we conclude that one is a likely RR$ab$ and that the classifier has thus only one false positives.

All in all the harmonic mean of false positives and false negatives is 1.71 or $\approx 4.4\%$, consistent with the cross-validation estimate and even better than expected.

\subsection{RR$ab$ in the Outer Bulge Area of the VVV}

\begin{figure}
\includegraphics[width=0.4\textwidth]{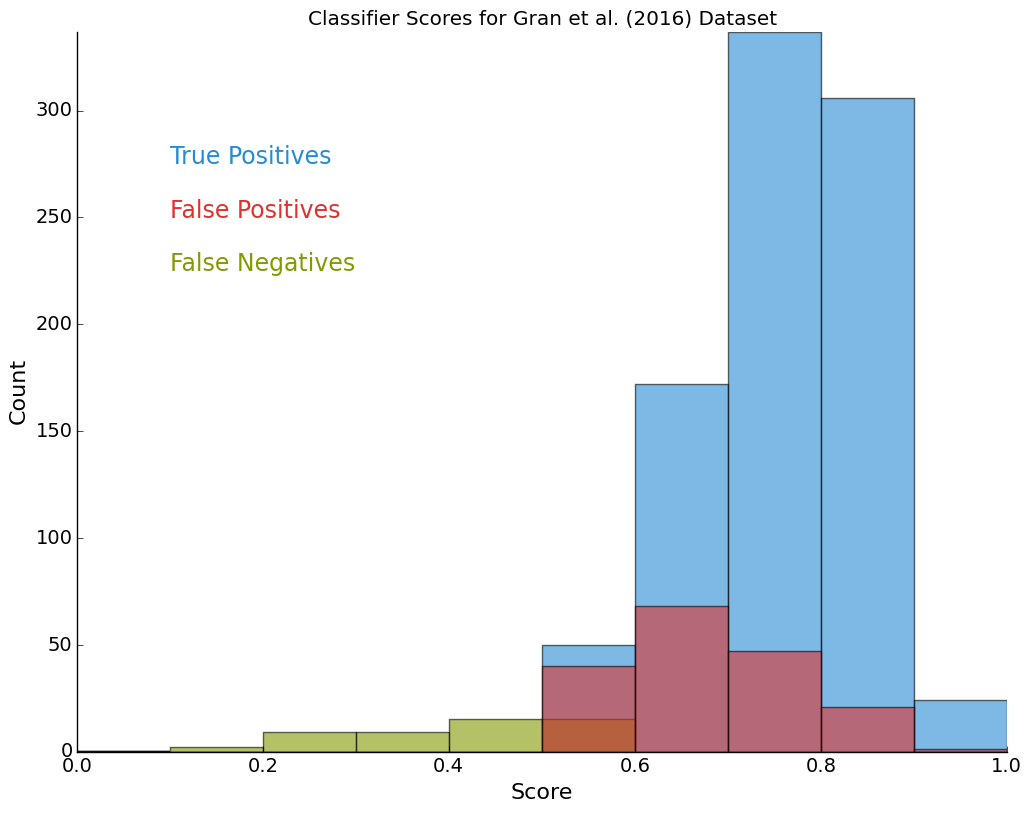}
\caption{Histogram of scores obtained by the classifier for the outer bulge light-curves of the Gran et~al.\ (2016) sample. Shown are the true positives (sources classified by as RR$ab$), false positives and false negatives.}
\label{fig:outer_scores}
\end{figure}

\citet{gran:2016} analyzed 7869 light curves consistent with being variable in the outer bulge region of the VVV survey and performed a human classification into RR$ab$ or ``other". The classifier presented in this work was run on the same dataset, and its results contrasted with the human expert performance. There were 1019 light curves classified as RR$ab$ in Gran et~al., of which 939 pass the cleaning process detailed in \S~\ref{ssec:outlier}. All the sources that were formally false positives and false negatives in comparison with Gran et~al. (2016) were re-assessed by eye; this was necessary given that our processing of the light curves was different and in particular the periods were estimated in a different fashion, leading in some cases to {\em bona-fide} RR$ab$ in our dataset that were not catalogued in Gran et~al. We found that among the 939 sources there were 50 false negatives, and the classifier gave an additional 177 false positives giving therefore a total of 1066 sources deemed as RR$ab$ by the classifier.
The distributions of scores for the outer bulge light curves is shown in Figure~\ref{fig:outer_scores}.
 The harmonic mean of the number of false positives and negatives is $\approx$78, or $\approx 8\%$ of the sample size, slightly lower but fully consistent with the $F_1$ measure as estimated from cross-validation on the training set.

We conclude that independent datasets that were not used in the training fully confirm the performance estimates obtained using cross-validation of the training sets.

%%%%%%%%%%%%%%%%%%%%%%%%%%%%%%%%%
%
% Discussion
% 
%%%%%%%%%%%%%%%%%%%%%%%%%%%%%%%%%

\section{Summary and Conclusions}
\label{sec:conclusions}

In this work we have presented the construction of a machine learned RR Lyrae type $ab$ classifier for the near infra-red light curves arising from the VVV survey. After preprocessing light-curves based on number of observations and outliers based on standard errors, we search for an  appropriate training set and the best aperture for each light-curve to feed the classifier. 
We look for the best supervised classifier among a wide variety of algorithms and found AdaBoost classifiers to consistently achieve the best performance. It is interesting to note that an AdaBoost classifier performed slightly above Random Forests, which have been gaining much popularity for classification problems in Astronomy, and in particular they have seen wide use recently in the problem of variable star classification \citep[e.g.][]{dubath:2011, richards:2012:asas, armstrong:2016}. Similar to Random Forest, the AdaBoost classifier is an ensemble method. These types of methods combine scores from several ``weak" classifiers to obtain better predictive performance. The key attribute of the AdaBoost is that it is adaptive  in such a way that samples that have been incorrectly classified by the previous ``weak" classifiers get higher weights in subsequent classifiers. Thus, the classifier adaptively concentrates on the training instances that are more difficult to classify.
In addition to being a slightly better performer, the AdaBoost classifier performance was stable with respect to sign all features or a restricted set of the most important features. Such stability is welcome as it results in a more forgiving behaviour with respect to the set of features chosen for a final classifier as long as there are enough features to capture the bulk of the classification problem.
We thus recommend that AdaBoost always be tried as an alternative, and in particular for the problem of variable star classification.

We extracted $68$ features  for each light-curve, and found that  $12$ were enough to achieve the best performance. The most important feature is the period, as expected due to the very well defined period range of RR$ab$ variables. Various phases and amplitudes of a harmonic fit to the light curve are also important in defining the RR$ab$ class. The performance of the chosen classifier, the AdaBoost.M1 algorithm \citep{freund:1996}, with the 12 most important features reaches an $F_1$ measure of $\approx 0.93$. This performance was estimated using cross-validation on the training sets, and was verified on two completely independent sets that were classified by human experts. 

The performance achieved by the classifier constructed in this work will allow us to tackle one of the main aims of the VVV survey, namely the identification of RR$ab$ over the full survey area with an automated, reproducible and quantitative classification tool. The expected harmonic mean between false positives and false negatives is of order $\approx 7\%$ when using a score threshold of 0.548 which maximizes this measure. In case purer or more complete samples are needed for particular science aims, the threshold on the classifier score can be adjusted as the needs dictate. For example, if a very complete {\em and} pure sample is desired, one could use a lower score threshold than the one that maximizes the $F_1$ score, and then further cull the list using ancillary information. The benefit of the classifier in this case is to reduce the number of candidate sources in an efficient, homogenous and reproducible way. 

Ancillary data to further assess the nature of variable light curves could come from the VVV itself, as there is also color information that is not being used by the classifier. The latter fact is by design, to avoid the complications and potential spatial biases that could be introduced in a classification scheme that uses colors. Given that analyzing the spatial structure of the Milky Way bulge is one of the main aims of the VVV, we want the classification procedure to be as free of spatial biases as possible. Of course, redenning will introduce spatial variations in the depth of the sample, but the classification should perform similarly for light curves of similar signal-to-noise ratio and phase sampling.

There are many more variables of interest in the VVV besides RR$ab$ \citep[see, e.g.][]{catelan:2013}. Future work will address the construction of machine-learned classifiers for those. We aim to design both custom binary classifiers for sources of interest, such as done in this work, but also a single multi-class classifier similar to that of \citet{richards:2012:asas}. We also plan to explore alternative classifier schemes beyond the supervised classification scheme employed here; an interesting possibility is to explore the use of Kohonen self-organizing maps, which has been tried before for the problem of the classification of light curves \citep{brett:2004,armstrong:2016}. It will be interesting to see if alternative classifier schemes can provide an improvement over the supervised scheme we use in this work, which already achieves a very good performance which is close to the performance of similar clasifiers in the optical where classification of RR$ab$ is easier.

\begin{acknowledgements}

S.E. and A.J. acknowledge useful discussions with Branimir Sesar.
This work was partly performed at the Aspen Center for Physics, which is supported by National Science Foundation grant PHY-1066293.
This work was partially supported by a grant from the Simons Foundation.
We acknowledge support by the Ministry for the Economy, Development, and Tourism's Programa Iniciativa Cient\'{i}fica Milenio through grant IC\,120009, awarded to the Millennium Institute of Astrophysics (MAS).
Ministry of Economy. 
JA-G acknowledges support from the FIC-R Fund, 
allocated to project 30321072, from FONDECYT Iniciaci\'on 11150916, 
and from CONICYT's PCI program through grant DPI20140066
Support by Proyecto Basal PFB-06/2007, FONDECYT Regular 1141141, and CONICYT/PCI DPI20140066 is also gratefully acknowledged.
G.H. acknowledges support from CONICYT-PCHA (Doctorado Nacional
2014-63140099) and CONICYT Anillo ACT 1101. N.E. is supported by CONICYT-PCHA/Doctorado Nacional.
R.K.S. acknowledges support from CNPq/Brazil through
projects 310636/2013-2 and 481468/2013-7.
F.E. acknowledges support from CONICYT-PCHA (Doctorado Nacional 2014-21140566). D. M. acknowledges support from FONDECYT Regular 1130196.

\end{acknowledgements}

\bibliographystyle{apj}
\bibliography{rrlyr.bib}

\end{document}